# EXTREMELY LONG C-C BONDS PREDICTED BEYOND 2.0 Å

Eero J. J. Korpela[1], Jhonatas Carvalho[2], Hans Lischka[2], Miklos Kertesz*[1]

[1] Chemistry Department and Institute of Soft Matter, Georgetown University
37th and O Streets, NW, Washington, DC 20057-1227, USA

[2] Department of Chemistry and Biochemistry, Texas Tech University, Lubbock, TX 79409, USA

*Email: kertesz@georgetown.edu

**Abstract**

A number of conjugated molecules are designed with extremely long single C-C bonds beyond 2.0 Å. Some of the investigated molecules are based on analogs to the recently discovered molecule by Kubo et al. These bonds are analyzed by a variety of indices in addition to their equilibrium bond length including the Wiberg bond index, bond dissociation energy (BDE), and measures of diradicaloid character. All unrestricted DFT calculations indicate no diradical character supported by high-level multireference calculations. Finally, $N_{FOD}$ was computed through fractional orbital density (FOD) calculations and used to compare relative differences of diradicaloid character across twisted molecules without central C-C bonding and those with extremely elongated C-C bonds using a comparison with the C-C bond breaking in ethane. No example of direct C-C bonds beyond 2.4 Å are seen in the computational modeling, however, extremely stretched C-C bonds in the vicinity of 2.2 Å are predicted to be achievable with a BDE of 15-25 kcal/mol.

**Introduction**

A recent result by Kubo et al.[1] showed the presence of a chemical bond between two carbon atoms at $D_{CC}$=2.042 Å, as identified by X-ray diffraction (XRD) in a highly strained environment. This remarkable finding was realized by two perpendicularly facing fluorenyl rings in the tris(9-fluorenylidene)methane, **1A**, a kind of butterfly shape with the two "wings" being joined at the "body" illustrated in **Scheme 1** together with selected examples of extremely long C-C single bonds. The purpose of this work is to explore variations on this molecule computationally by looking for two questions: (i) Is it possible to obtain molecular structures with even longer single bonds? (ii) What are the special features of these extremely long single bonds?

In this article, we place this discovery by Kubo et al. in the context of the historical progression of longer and longer single bonds obtained in several laboratories over the years, all of which displayed bond lengths as long as the recent 1.93 Å value for a diamino-o-carborane[2] following on the heels of others at 1.8 Å[3], at 1.77 Å[4], and around 1.7 Å somewhat earlier.[5, 6, 7, 8, 9] The example of a carborane contains a C-C distance between two six-coordinated carbons as long as 1.93 Å[10] and another with 1.99 Å,[11] with a Wiberg bond index of 0.33. Mandal and Datta describe carborenes with C-C bonds as long as 2.01 Å.[12] During these series of discoveries as the limit of the longest C-C single bond has been gradually pushed to larger and larger values. Ishigaki et al. reasoned a few years ago that molecular examples with C-C bonds longer than 1.8-2.0 Å should be forthcoming.[3] **Scheme 1** displays some of these molecules with unusually long C-C bonds in organic molecules. Organic ligands in transition metal complexes occasionally also display very long single C-C bonds, e.g. by Han et al.[13] at 1.87(2) Å.

There are no unambiguous theoretical reasons why the longest two-electron single bond between two carbon atoms must break at about 2.0 Å. Alvarez has surveyed the periodic table searching for improved van der Waals radii and for the presence or absence of a "van der Waals gap" in the distribution of contact distances in the CSD and finds one for the carbon atoms bound to an oxygen atom.[14] On the theory side, based on atoms-in-molecules and electron localization function computations Isea argued that C-C single bonds should still show key characteristics of sigma bonds up to approximately 2.0 Å, but not beyond 2.0 Å.[15] Based on the analysis of a large database Lobato et al. arrived at a similar conclusion recently.[16] It is interesting that as Kubo et al.[1] noted, based on careful temperature-dependent XRD analysis, that the intrinsic distance of the long C-C bond in **1** is somewhat shorter than 2.042 Å close to ~1.98 Å due to crystal packing effects. Cho et al.[17] have argued that this limit should be about 1.8 Å, slightly longer than suggested previously by Zavitsas[18] and Schreiner et al.[8] based on the dependency of the binding energy as a function of the long C-C bond distance between two sp$^3$ carbon atoms connected to adamantanes or alkanes. They estimated that at about 1.8 Å the C-C bond dissociation energy (BDE) becomes very small or zero in the series of highly crowded adamantanes. While the accurate assessment of the BDE is challenging, its value is of importance in the presented discussions as we evaluate its approximate value with a singlet-triplet energy gap. Overall, based on the history of the problem, any C-C bond distance longer than 1.8 Å should be considered unusual and worthy of analysis.

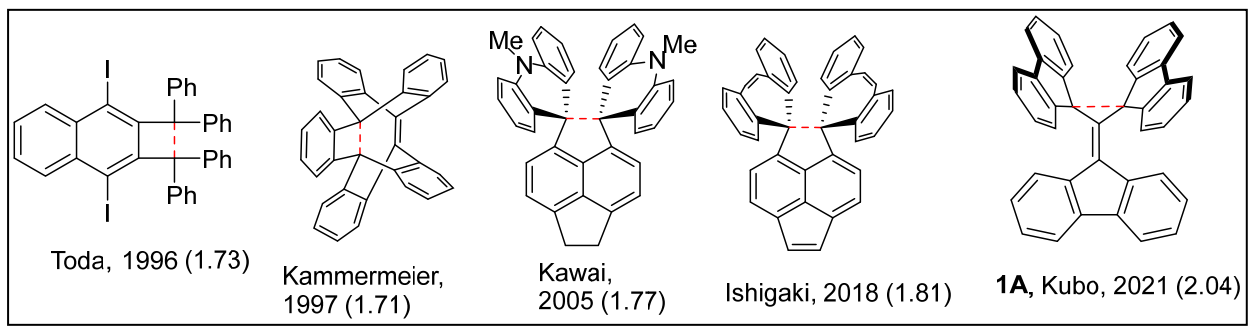

**Scheme 1**. Selected experimentally characterized examples of very long C-C single bonds. Numbers in parenthesis following the first author's name and year are in Å and indicate the length of the long bond shown by a red dashed line.

Based on the discovery by Kubo et al. and previous cases of very long C-C single bonds, we have continued to ask the questions whether (i) examples can be found where even longer bond lengths, and (ii) whether these elongated bonds still display main characteristics of a C-C single chemical bond.

The first question (i) can be addressed in a relatively straightforward manner by investigating the equilibrium geometries of proposed molecules with computational methods that are sufficiently reliable in predicting geometries and relative energies. The second question (ii) is more subtle. There are a number of physical parameters that can be used to characterize and compare the strengths of chemical bonds; none of them perfect, especially when applied to weak bonds. Another complication is that in many weak and long single bonds, steric repulsion plays a significant role[19], as if the effect would be primarily due to bond stretching. In many of these cases, the separation of these opposing effects, bond formation and steric repulsion leading to bond stretching, is to some degree elusive and arbitrary.

**Scheme 2** indicates an intriguing feature of the long C-C bond in **1**: a simple VB argument would indicate some, possibly strong, diradicaloid character, as is the case with highly stretched bonds.[20] Notwithstanding, Kubo et al.[1] convincingly argue based on CASSCF(6,6)/6-311G(d) computations that molecule **1A** has a very small diradicaloid character as measured by the $y_0$ index of 0.128 in the ground state of this molecule.

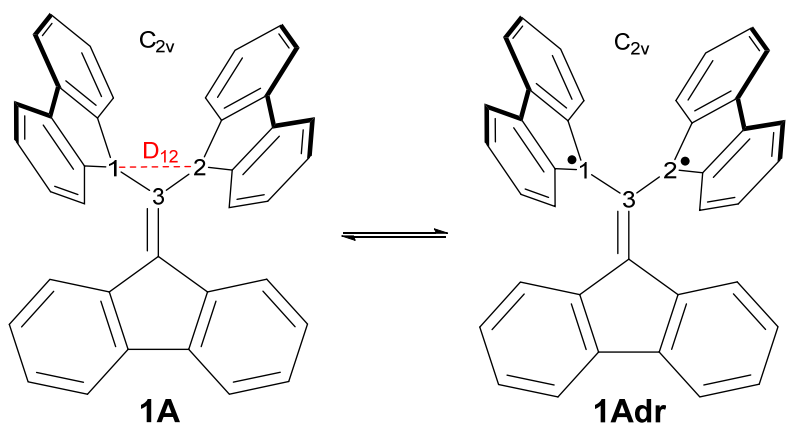

**Scheme 2**. Two VB structures of **1A** (covalent) and **1Adr** (diradical). Note the $C_{2v}$ symmetry for both VB structures of the isolated molecule, **1A**, as found in the crystal structure.[1]

An additional complication in investigating extremely long and therefore relatively weak covalent single C-C bonds is the possibility that the bond can break resulting in a diradical isomer. This possibility is present, for example, for **2A** in the form of a twisting deformation, as illustrated in **Scheme 3**. It will be interesting to explore these deformations, the energetics of these isomerization reactions, and how to prevent them should they lead to a lower energy twisted diradical, which in fact turns out to be the case in more than one of the presented results.

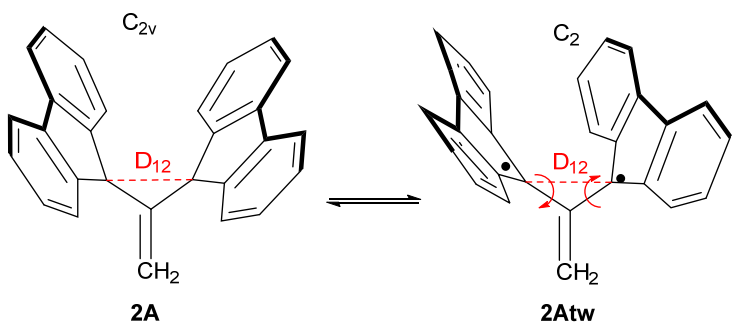

**Scheme 3**. Isomerization reaction involving twisting of the "wings" of some of the molecules discussed. Red arrows indicate the conrotatory twists. **2Atw** is a structural isomer that has a local minimum on the computed potential energy surface.

In the following, we will characterize the very long covalent single C-C bonds, identified as $D_{12}$, using accessible parameters in addition to the equilibrium bond distance ($R_e$), including the Wiberg bond index (WBI)[21], and the bond dissociation energy (BDE). In addition, we will be interrogating these weak bonds by their diradical character, which serves to indicate a measure of the degree of dissociation and the degree of electron pairing in the bond. The discussion of the diradicaloid character of extremely stretched bonds has been a common theme in most studies[1, 3, 20] as a way to describe how far along the dissociation a particular stretched bond may be. We generally found a low level of diradicaloid character for bond distances up to even 2.0 Å. A further measure of the strength of the covalent bonds investigated is provided by the singlet-triplet energy difference ($\Delta E_{ST}$), which becomes small as the bond approaches dissociation.[22]

Before enumerating the methodology and turning to the results, one comment on terminology can be helpful to avoid a possible misunderstanding. There is a category of weak C-C bonds, typically binding radicals together that are characterized by multicenter electron sharing, the prototypical example being the pairing of phenalenyl (PLY) dimers. The C…C contact distances in these so-called pancake bonds are shorter than twice the van der Waals radius of carbon at $D_{vdW}$=3.40 Å.[23] The shortest of these observed by XRD was for a dimer of tetracyanoethylene anion radical (TCNE$^{-2}$)$_2$ at 2.801 Å.[24] However, these pancake bonds, due to their multicenter nature, e.g. a two-electron 12-center (2e/12c) bond for PLY$_2$, and a two-electron 4-center (2e/4c) bond for (TCNE$^{-2}$)$_2$, should not be compared with the long single bonds in molecules shown in **Scheme 1**, or their analogs discussed here.

A further distinction relates to fluxional bonding. While the focus in this work is on very long equilibrium bond distances ($R_e$), XRD data may indicate extremely long C-C bond distances that correspond in fact to an average of a bond distance distribution due to fluxional bonding, as they may occur for example in bisnorcaradienes with $R_{XRD}$ as long as 1.8 Å.[25, 26, 27] In the crystal structure of dimers of phenalenyl derivatives the observed $R_{XRD}$=2.153 Å[28] is the result of large amplitude fluxional bonding, not to be confused by equilibrium bond distances.[29, 30]

The target region of C-C bonds discussed in this paper is indicated by a green rectangle in **Figure 1**, which is larger than the typical stretched single C-C bond and shorter than pancake bonds.

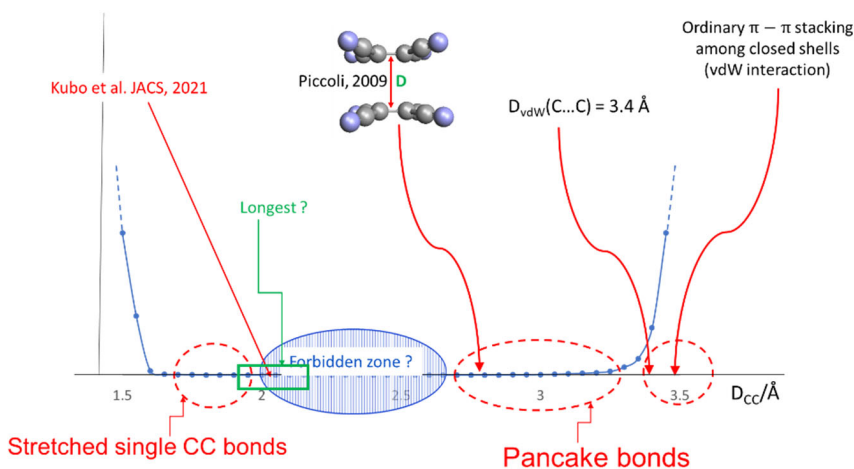

**Figure 1.** Schematic representation of a hypothetical histogram of unusual carbon-carbon bonds and contacts ($D_{CC}$). Molecules discussed in this work are encroaching on the "forbidden zone"[14] from the left

and correspond to bond distances represented by the green rectangle. Blue points and line indicate only the rapid rise of the relative numbers of such contact distances on the left and on the right, hence no specific vertical scale is indicated emphasizing the relative scarcity of extremely stretched bonds and pancake bonds.

**Methods**

Full geometry optimizations yielding the equilibrium stretched C-C bond length ($R_e$) have been performed with UB3LYP level of density functional theory (DFT) with empirical dispersion term included in the total energy using the GD3 parametrization[31], where U indicates the spin unrestricted version. The 6-311+G(d,p) basis set was used except where noted otherwise. Each local minimum or transition structure (TS) was confirmed by zero or one imaginary frequency, respectively. To investigate the diradicaloid character of the electronic structure of molecules, UB3LYP/6-311+G(d,p) calculations were run for all molecules while higher level multireference-averaged quadratic coupled cluster[32] (MR-AQCC) calculations were run for ethane. Several descriptors were used to characterize the diradicaloid character of a molecule. The $y_0$ parameter as a descriptor of diradicaloid character was calculated according to the formula[33]

$$y_0 = \text{NOON}_{LU} \tag{1}$$

where $\text{NOON}_{LU}$ is the natural orbital occupation number (NOON) for the lowest unoccupied orbital.[34] In addition to the $y_0$ parameter, Fractional Orbital Density[35,36] (FOD) calculations B3LYP/6-311G+(d,p), with the recommended electronic Fermi temperature of $T_e$=9000 K, were completed as another measure of diradicaloid character. Note that all FOD computations refer to the restricted DFT. The FOD analysis provides a quick measure of diradical character through $N_{FOD}$, a parameter obtained by spatial integration of the FOD. To further probe the accuracy of the FOD analysis as a measure of diradical character, MR-AQCC/6-311G+(d,p) calculations were performed for the dissociation of ethane. The reference wavefunction, which was also used for initial multiconfiguration self-consistent field calculations, was constructed within a general valence bond (GVB) perfect-pairing multiconfigurational (PPMC) approach.[37,38] This wavefunction is of direct-product form where electron pairs are assigned to pairs of active orbitals whose occupancies are determined variationally. Only singlet coupling of all electron pairs were allowed. These MR-AQCC calculations were used to obtain the potential energy curve and the number of effective unpaired electrons, $N_U$, in the relaxed dissociation of ethane. The $N_U$ values were obtained according to the nonlinear formula of Head-Gordon[39] as

$$N_U = \sum_i n_i^2 (2 - n_i)^2, \tag{2}$$

where $n_i$ is the occupation of the $i^{th}$ natural orbital (NO) and the sum is over all NOs. **Figure 2** shows the MR-AQCC dissociation curve and the evolution of $N_U$ with increasing C-C distance. For comparison the FT-RDFT/B3LYP dissociation curve and $N_{FOD}$ values calculated with the FT-RDFT/B3LYP/6-311+G(d,p) method are also shown. Both methods produce almost identical potential energy curves. Similarly, the $N_{FOD}$ values are well described with the FT-RDFT method with values ~2 $e$ in the dissociation region. This behavior coupled with the observation that $N_{FOD}$ values correlate well with $N_U$ values obtained at the MR-AQCC level indicates that the fractional occupation is well reproduced by the FT-RDFT method. Based on these results, $N_{FOD}$ was used to compare the diradical character across all molecules included in the study. The Gaussian 16 program was used in most of this work. For the FOD calculations the ORCA 5.0 program was used.[40,41] The MR-AQCC calculations were performed with COLUMBUS.[42,43]

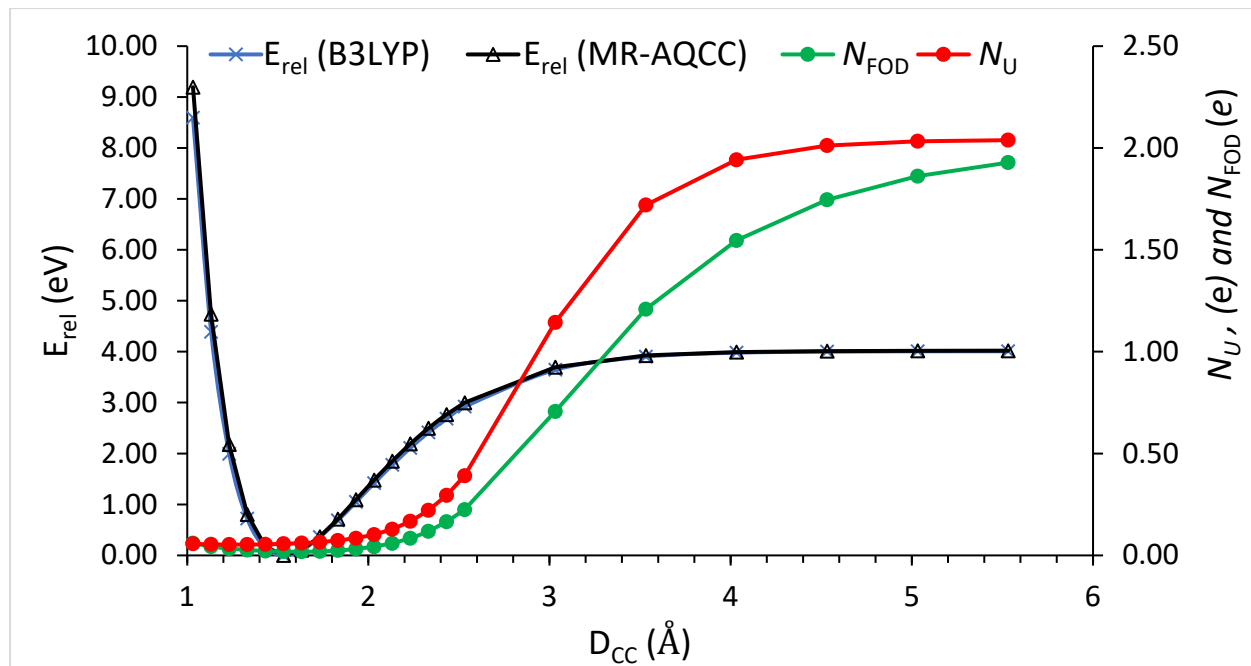

**Figure 2.** Potential energy curves (in relation to the minimum geometry) for relaxed displacement along the C-C bond in ethane and $N_U$ values calculated with the MR-AQCC(PPMC)/6-311+G(d,p) and $N_{FOD}$ FT-RDFT/B3LYP/6-311G+(d,p) methods.

The evaluation of the bond strength via BDE is essential. Unfortunately, in the systems under consideration, a simple dissociation of the highly stretched C-C bond is not possible due to the complex topology of these molecules that engender various strains and tethering. Consequently, we have employed two indirect approaches to estimate the BDE of the long C-C bonds. Here we summarize these computation protocols and their justifications. It needs to be noted that the separation of strain and other relaxation from the intrinsic BDE is not trivial and is by definition model dependent. Nevertheless, we expect that useful trends will emerge from these data and their comparisons.

(1) Estimation of BDE by considering the vertical transition from the singlet ground state to the triplet excited state, according to this formula:

$$BDE_{ST} = \Delta E_{S-T} = E_S(singlet\ optimized\ geometry) - E_T(triplet\ single\ point\ calculation) \quad (3)$$

Here $E_S$ and $E_T$ are the singlet ground state and lowest triplet state energies, respectively, computed by the spin-unrestricted formalism.

The approximation of the BDE in this manner goes back to the analysis of single bond dissociation by Michl.[44] Similar approaches have been used for other weakly bonded systems.[45,46] Kubo et al. estimated the respective BDE for **1A** to be 138 kJ/mol (33 kcal/mol). In a relaxed geometry version of the same method where the triplet geometry was also optimized, they obtained a BDE of 113 kJ/mol (27 kcal/mol) for **1A**, noting that due to relaxation, this measure includes the release of some of the angle strain seen in **1A**.[1]

(2) A second approach relies on the possible presence on the potential energy surface (PES) of an isomeric structure without the weak bond in question. Such structures may be present in some cases, and not in others. In the cases where these structures are present, the $BDE_{isomers}$ refers to the energy difference between a non-twisted and twisted conformer, as seen in **Scheme 3**. The BDE obtained in this manner is the following:

$$BDE_{isomers} = E_S(bonded\ optimized) - E_S(unbonded\ optimized) \qquad (4)$$

The values for $BDE_{isomers}$ obtained in this manner can be strongly affected by the differences in the strain energies of the two isomers and therefore turn out to be less useful than $BDE_{ST}$.

$^{13}$C NMR calculations were run for selected target molecules, whereby their $^{13}$C NMR chemical shifts were computed by the GIAO-UB3LYP-GD3/6-311+G(d,p) method[47]. These structures were optimized using the same level of theory. TSM was used as the reference computed also using GIAO-UB3LYP-GD3/6-311+G(d,p).

**Molecular Design**

For all the target molecules of this work, the two carbon atoms in question have three other carbon atoms attached to them in addition to the long bond being investigated. In this sense, they are analogues of the molecules shown in **Scheme 1**. All target molecules in this study can be seen in **Table 1**. Moreover, the names of each molecule are defined using **Table 2**, where the first column represents the "body" and second column the "wings" of these butterfly shaped molecules.

**Table 1.** List of target molecules in their non-twisted configurations investigated in this study.

| | | | | | | | |
|---|---|---|---|---|---|---|---|
| 1A | 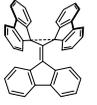 | 1A (Me:4,8) | 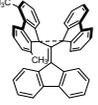 | 2A | 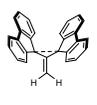 | 3A | 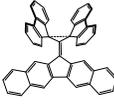 |
| 1A (CN:4,12) | 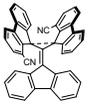 | 1A (Me:4,9) | 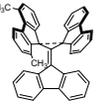 | 2A (F:4,12) | 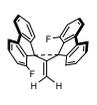 | 5F | 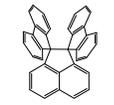 |
| 1A (F:4) | 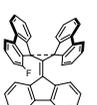 | 1A (Me:4,10) | 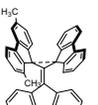 | 2A (Me:4) | 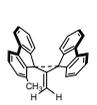 | 6F | 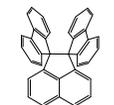 |
| 1A (Me:4) | 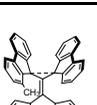 | 1A (Me:4,11) | 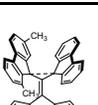 | 2A (Me:4,12) | 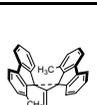 | 7F | 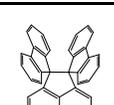 |
| 1A (Me:5) | 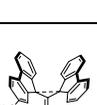 | 1A (Me:4,12) | 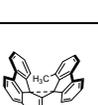 | 2A (Me:4,11,12,19) | 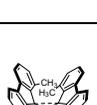 | 8F | 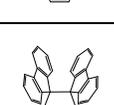 |
| 1A (Me:6) | 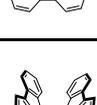 | 1A (Me:4,19) | 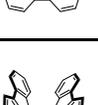 | 2A (Br:4,11,12,19) | 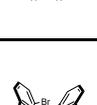 | 9F | 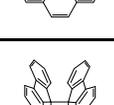 |
| 1A (Me:7) | 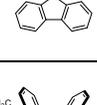 | 1A (OH:5,10,13,18) | 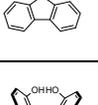 | 2B | 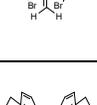 | 10A | 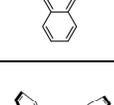 |
| 1A (Me:4,5) | 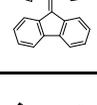 | 1B | 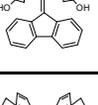 | 2C | 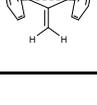 | 10A (Me:4) | 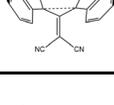 |
| 1A (Me:4,6) | 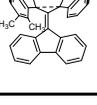 | 1C | 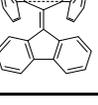 | 2D | 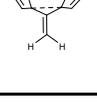 | 10D | 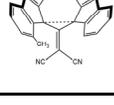 |
| 1A (Me:4,7) | 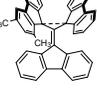 | 1D | 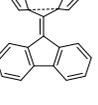 | 2E | 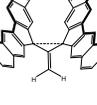 | 11A | 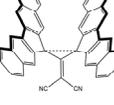 |

**Table 2.** Key for target molecules within this study. Each target molecule corresponds to a number identifying the "body" and a letter for the "wings" possibly with additional substituents. The red highlighted bonds indicate the linking units for the "wings" and "body". E.g., **1A** corresponds to the molecule in **Scheme 2**, and **2A** corresponds to the one in **Scheme 3**.

| | | | |
|---|---|---|---|
| 1 | 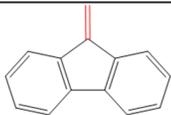 | A | 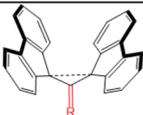 |
| 2 | 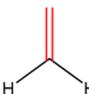 | B | 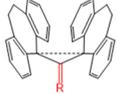 |
| 3 | 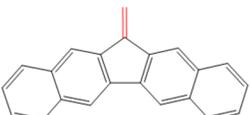 | C | 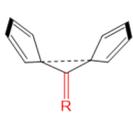 |
| 4 | 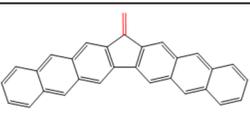 | D | 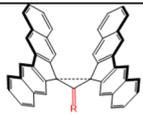 |
| 5 | 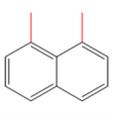 | E | 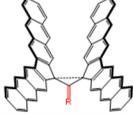 |
| 6 | 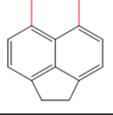 | F | 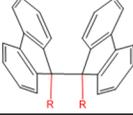 |
| 7 | 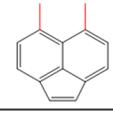 | | |
| 8 | 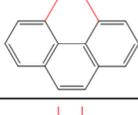 | | |
| 9 | 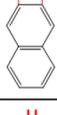 | | |
| 10 | 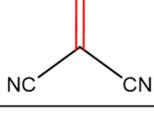 | | |
| 11 | 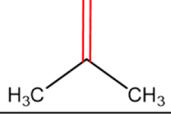 | | |

Molecules with letter codes B-E are related to and derived from those with a letter code A. Their distinguishing feature is the number of rings and maximum ring size of their "wings". Similarly, molecules with a number code 2 and 3 are related to molecules with the number code 1, except each number corresponds to different a "body", as seen in **Table 2**. Molecules 5F-8F are related to and derived from Ishigaki's molecules, one of which is seen in **Scheme 1**. Molecule 9F is related to the Toda molecule in **Scheme 1**. Finally, molecules 10A and 11A are derivatives of **2A** with electron-withdrawing or electron-donating groups on all available sites of the molecule's "body". For some of these molecules, there are other derivatives with various substituents on their "body" and "wings" that were included as target molecules. **Scheme 4** shows atomic numbering used in this work.

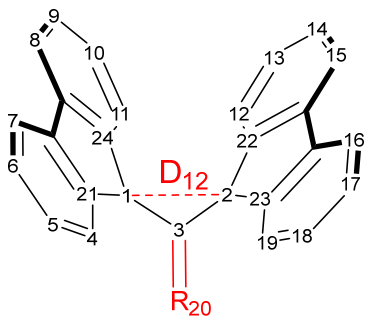

**Scheme 4.** Numbering used to identify target molecules with additional substituents on their letter code A "wings". E.g., **1A(Me:4)** corresponds to molecule 1A with a methyl substituent at C4.

**Results and Discussion**

Key results of the computational modeling are presented in **Table 3**. The results consistently indicate extremely long covalent single C-C bonds with equilibrium bond distance $R_e$ values in the range of 1.6-2.2 Å, some of which are remarkably long, placing them in the unusual category within the green rectangle in **Figure 1**. The molecules that stand out having the longest $R_e$ values will be discussed.

**Table 3.** Key results for non-twisted target molecules from computational modeling calculated at the UB3LYP-GD3/6-311+G(d,p) level of theory.

| Molecule | $R_e$ (Å) | WBI | $BDE_{ST}$ (kcal/mol) | $N_{FOD}$[a] [e] |
|---|---|---|---|---|
| 1A | 2.048 | 0.437 | 32.4 | 1.76 |
| 1A(CN:4,12) | 2.071 | 0.409 | 28.2 | 1.98 |
| 1A(F:4) | 2.071 | 0.431 | 28.7 | 1.82 |
| 1A(Me:4) | 2.151 | 0.354 | 19.8 | 2.03 |
| 1A(Me:5) | 2.052 | 0.433 | 20.9 | 1.79 |
| 1A(Me:6) | 2.068 | 0.421 | 30.2 | 1.81 |
| 1A(Me:7) | 2.073 | 0.417 | 29.7 | 1.82 |
| 1A(Me:4,5) | 2.118 | 0.379 | 24.9 | 1.94 |
| 1A(Me:4,6) | 2.165 | 0.341 | 18.3 | 2.08 |
| 1A(Me:4,7) | 2.157 | 0.347 | 19.3 | 2.06 |
| 1A(Me:4,8) | 2.152 | 0.342 | 20.0 | 2.05 |
| 1A(Me:4,9) | 2.153 | 0.352 | 19.6 | 2.05 |
| 1A(Me:4,10) | 2.137 | 0.363 | 22.0 | 2.01 |
| 1A(Me:4,11) | 2.097 | 0.405 | 27.7 | 1.89 |
| 1A(Me:4,12) | 2.191 | 0.321 | 15.4 | 2.16 |
| 1A(Me:4,19) | 2.117 | 0.390 | 25.5 | 1.93 |
| 1A(OH:5,10,13,18) | 2.070 | 0.416 | 32.3 | 1.99 |
| 1B | 1.641 | 0.801 | 63.7 | 1.46 |
| 1C | 1.625 | 0.765 | 68.9 | 0.88 |
| 2A | 2.099 | 0.388 | 27.8 | 1.37 |
| 2A(F:4,12) | 2.107 | 0.343 | 24.7 | 1.43 |
| 2A(Me:4) | 2.144 | 0.400 | 22.7 | 1.50 |
| 2A(Me:4,12) | 2.192 | 0.399 | 17.1 | 1.65 |
| 2A(Me:4,11,12,19) | 2.180 | 0.379 | 21.2 | 1.59 |
| 2A(Br:4,11,12,19) | 2.170 | 0.368 | 20.1 | 1.78 |
| 2B | 1.639 | 0.802 | 62.9 | 1.00 |
| 2C | 1.636 | 0.763 | 67.2 | 0.43 |
| 2D | 2.183 | 0.312 | 23.1 | 2.14 |
| 2E | 2.228 | 0.247 | 23.4 | 3.18 |
| 3A | 2.049 | 0.436 | 32.3 | 2.13 |
| 5F | 1.636 | 0.868 | 83.1 | 1.02 |
| 6F | 1.664 | 0.883 | 76.8 | 1.06 |
| 7F | 1.667 | 0.852 | 55.4 | 1.24 |
| 8F | 1.597 | 0.900 | 72.8 | 1.15 |
| 9F | 1.736 | 0.801 | 73.9 | 1.06 |
| 10A | 2.213 | 0.344 | 21.1 | 1.73 |
| 10A(Me:4) | 2.231 | 0.294 | 17.6 | 1.83 |
| 10D | 2.254 | 0.305 | 21.0 | 2.40 |
| 11A | 2.079 | 0.397 | 25.9 | 1.35 |

[a] $N_{FOD}$ calculations computed with B3LYP/def2-TZVP ($T_e$=9000 K) level of theory

While geometric parameters of typical C-C covalent bonds are nearly constant depending on orbital hybridization, throughout the literature there are several molecules such as those presented in **Scheme 1** that rely on steric effects to distort these typically stable geometric parameters under highly strained conditions.[3, 5, 6, 7] As a result, in this study, steric effects are a core strategy used to probe the limits of covalent single C-C bonds. One of the longest observed bonds in this study that utilizes steric hinderance as its primary mode of elongation is **1A(Me:4,12)** with an equilibrium bond distance of 2.191 Å. With methyl groups positioned at carbons 1 and 9, both fluorene "wings" are forced to separate from one another due to steric repulsion. This separation is not a simple elongation of the bond along the axis of where the bond exists, but rather a distortion of the geometry of this molecule by slight twisting of its "wings", adopting a $C_2$ geometry. This twisting is similar to that shown in **Scheme 3**, however, the molecule does not fully adopt a twisted conformer without an elongated central carbon bond, as confirmed by the zero diradical character, $y_0$, and relatively low $N_{FOD}$. Instead, this molecule twists slightly, which elongates the bond, to lower the van der Waals repulsion between the methyl substituents and hydrogen atoms on the opposing fluorenyl "wing". For most of the other molecules that have substituents on their "wings" and exhibit elongated bonds, a similar reasoning of steric hindrance can be used.

As explained above, the elongated bonds of the target molecules are too short to fall in the category of pancake bonds. However, for molecules with the large "wings" D and E, there appears to be pancake-like interactions between carbons between the two "wings", so a pancake bonding model[23] can be used to understand the attractive interaction between the "wings" in these systems. One such through space bonding interaction is indicated by the in-phase orbitals between the two "wings" seen in **Figure 3** for the HOMO of **2E**. The interaction is labeled between two carbons at a length of 2.922 Å, which is within the typical range of pancake bonding. If such pancake bonding would not be present, this short contact distance would imply large steric repulsion. A geometric consequence of this pancake-like interaction is reflected in the optimized geometries for these types of molecules with extended macrocycle "wings". Unlike for the fluorenyl "winged" molecules, these larger macrocycle molecules converged to energy minima where their "wings" almost completely eclipsed one another. This eclipsed conformation shortens the distance between wings, allowing for pancake-like bonding interactions. While this study did not focus on these interactions, it still should be noted that molecules with "wings" D and E have $D_{12}$ distances significantly elongated, surpassing many sterically hindered molecules. For example, **2E** has an equilibrium $D_{12}$ of 2.228 Å which is longer than all 1A and 2A sterically hindered molecules with $D_{12}$ distances ranging from 2.048 to 2.192 Å.

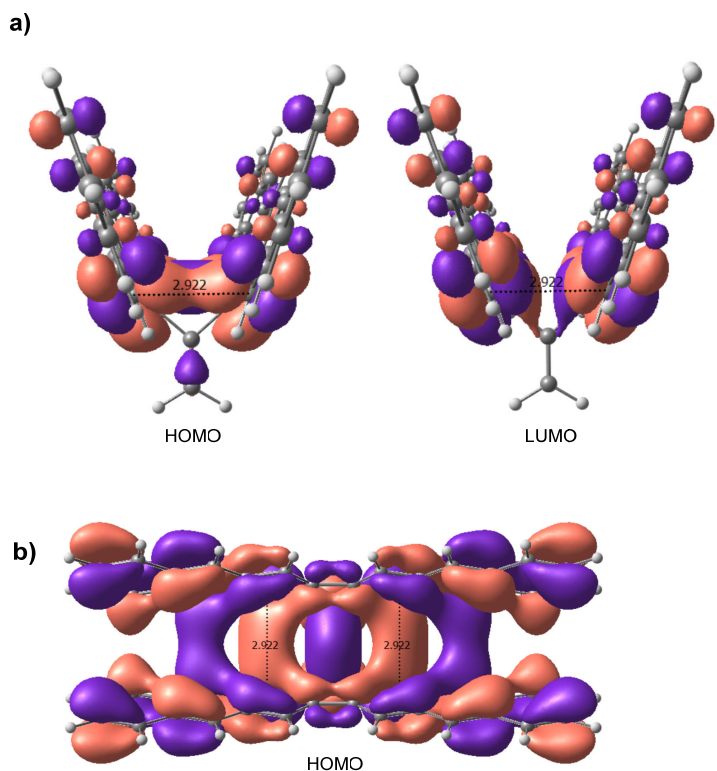

**Figure 3.** a) HOMO and LUMO of **2E** calculated using the B3LYP-GD3/6-311+G(d,p) method. Purple and red surfaces represent the relative signs of the orbital coefficients drawn at the 0.03 e au$^{-3}$ level. b) Alternative view of HOMO of **2E** drawn at a 0.01 e au$^{-3}$ level. The equilibrium distances indicated at 2.922 Å correspond to $D_{21,23}$ and $D_{22,24}$ in Scheme 4.

The effects of extending and shortening the "body" and "wings" of **1A** were investigated by two series of molecules: **1A**, **2A**, **3A** and **2A**, **2D**, **2E**. In the first series **1A-3A**, the fluorenyl body was shortened in **2A** to a methylene group and lengthened in **3A** to a 12H-dibenzo[b,h]fluorene group. The effect of altering the body of **1A** is unclear since both lengthening and shortening the body both had the effect of increasing $D_{12}$. However, the effect on $D_{12}$ was more pronounced when shortening the body to **2A** where $D_{12}$ increased by ~0.05 Å while lengthening the body marginally increased $D_{12}$ by less than 0.001 Å. However, because the body of all these Kubo-like molecules does not play a direct role in the bonding of $D_{12}$, as seen for example in the two frontier MOs of **2E** in **Figure 3**, it was expected that modifications of the "body" of **1A** would have little impact on $D_{12}$. In contrast, the effect on $D_{12}$ due to variations the "wings" was far more pronounced. Going from **2A** to **2D** to **2E,** the fluorenyl "wings" were extended on both sides by one benzene group, resulting in significant $D_{12}$ increases from 2.099 Å to 2.183 Å to 2.228 Å. These large increases in $D_{12}$ can be explained by the increase in pancake bonding as a result of larger macrocycle conjugated systems. Unlike in the first series where the "body" was systematically changed, the "wings" of the Kubo-like molecules play a large role in bonding. Specifically, C1 and C2, both part of the "wing" macrocycles, are the two carbons involved in $D_{12}$.

The inductive effect via electron-donating and electron-withdrawing substituents was also explored as a way to stabilize extremely elongated bonds. This effect was tested by adding electron-withdrawing and electron-donating groups to either the "body" or "wings" of **2A**. When adding electron-withdrawing and electron-donating groups to the "wings" of any molecule under study, steric effects dominated the observed response to bond length. For example, adding halogens or methyl groups to the "wings" resulted in partial twisting, a result of the confined space between wings leading to the elongation of the $D_{12}$ bond. In contrast, the effects of adding electron-withdrawing and electron-donating groups to the body of the target molecules were less clear. A decrease was seen in the bond length with the addition of electron-donating methyl groups to the body of 2A, **11A**. However, when adding electron-withdrawing groups to "bodies" of the target molecules, $D_{12}$ increased. Specifically, for the cyano substituted molecule (**10A**), a large increase in bond length was observed to 2.213 Å.

For all the target molecules in this study, correlations were constructed comparing bond length to a variety of parameters, including WBI, $BDE_{ST}$, and $N_{FOD}$. The respective WBI values correlate very well with $D_{12}$ as illustrated in **Figure 4**. These data indicate that several molecules in the dataset have significant bond orders with bond distances larger than 2.0 Å with WBI values of 0.3 and larger for bond distances of up to ~2.25 Å, which are significantly longer than that of the Kubo molecule. However, the nearly linear correlation indicates that no C-C WBI is expected beyond around 2.5 Å. It should be noted that molecules beyond 2.4 Å exhibit twisted conformers with significant diradical character and low WBI, indicating the absence of a bond.

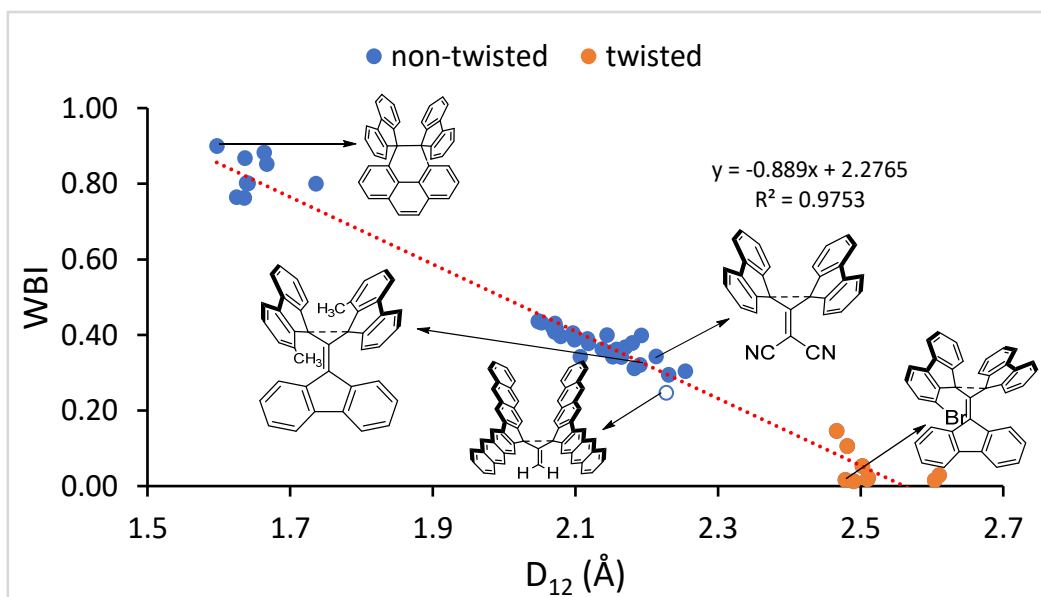

**Figure 4.** Correlation between $D_{12}$ and the WBI for molecules with extremely long covalent single C-C bonds. For selected data points, their corresponding molecules are shown. The non-filled blue data point refers to 2E.

    Bond dissociation energies are physically well-defined quantities compared to bond orders which are not.[48] However, as outlined in the methods section, a direct computation of the BDE in the presented cases is not possible. First, we display the $BDE_{ST}$ values in **Figure 5** that can be used as surrogates of the BDE as per equation (3). The trends are similar to that seen in **Figure 4** for the WBI except that the linear trendline indicates a shorter limit where the extremely stretched C-C bonding diminishes to the absence of any bonding at ~2.45 Å. The strength of the $BDE_{ST}$ computed in this manner becomes smaller than 10 kcal/mol at ~2.3 Å, which should be considered as the long limit of extremely stretched C-C bonds. However, molecule **10A** with the computed $R_e$= 2.213 Å still displays a significant $BDE_{ST}$ of 21.1 kcal/mol putting it on par with other very weak covalent bonds, such as the elongated bond (1.68 Å) present in 1,2-di(9-anthryl)benzene.[49] Thus molecules on this long limit of extremely stretched C-C bonds still display qualities of bonding character.

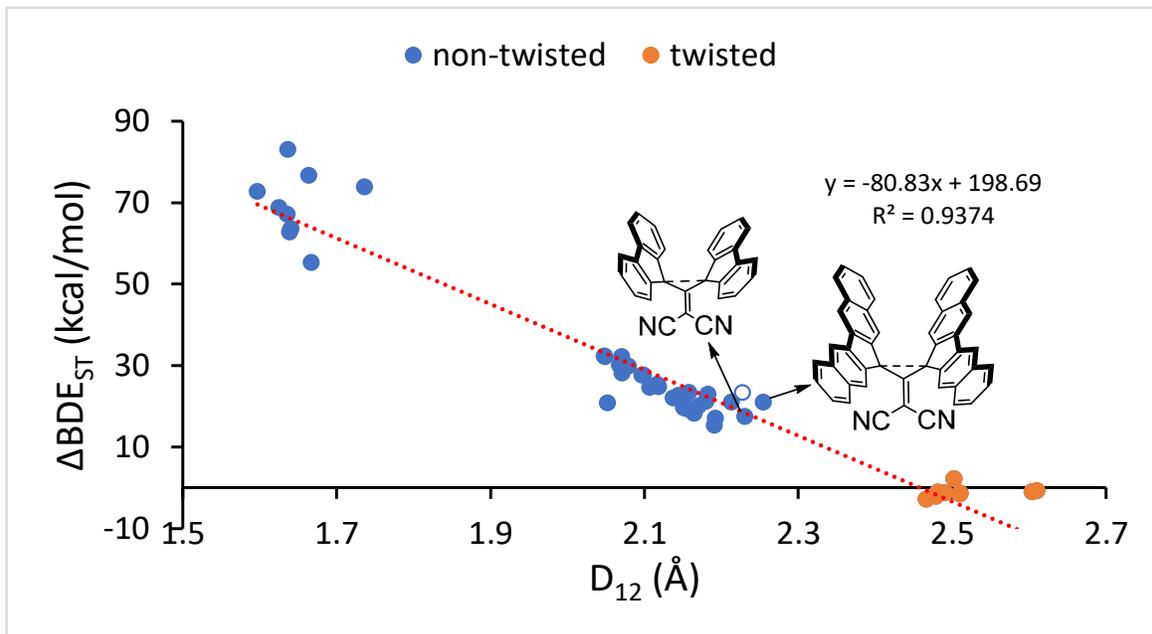

**Figure 5.** Correlation between $D_{12}$ and $BDE_{ST}$ for molecules with extremely long covalent single C-C bonds. The data points for molecules **10A and 10D** are indicated by arrows. The non-filled blue data point refers to 2E.

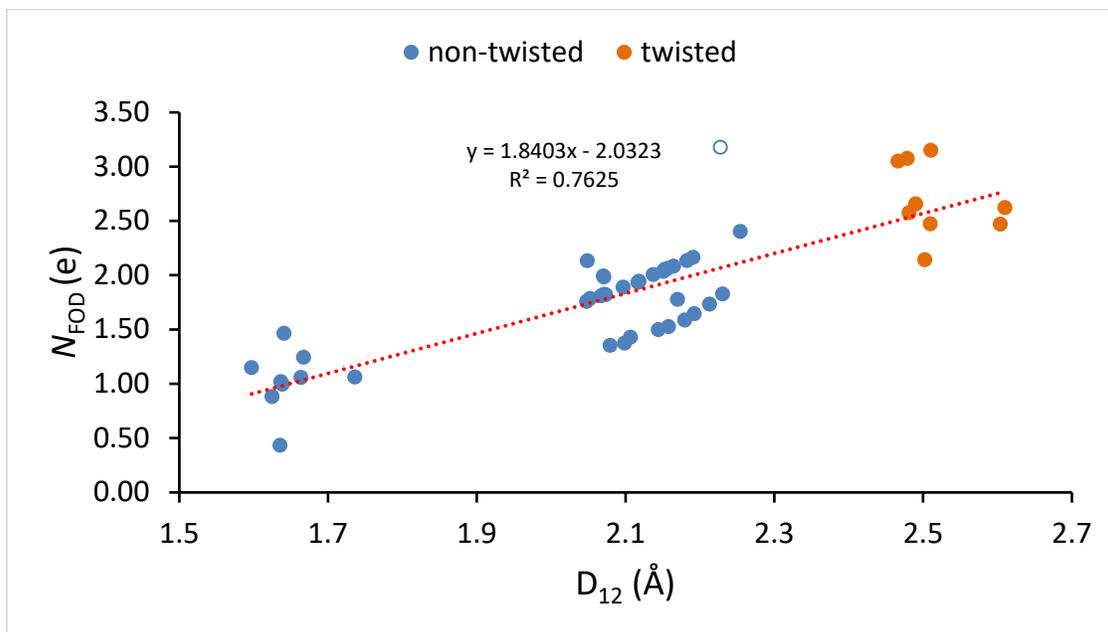

**Figure 6.** Correlation between $D_{12}$ and the $N_{FOD}$ for molecules with extremely long covalent single C-C bonds. The non-filled blue data point refers to **2E**, see text.

**Figure 6** illustrates the positive linear correlation between $N_{FOD}$ and $D_{12}$. While there is no absolute cutoff for $N_{FOD}$ that indicates the presence or absence of a C-C bond, it can be used as a relative measure of diradical character which increases as a covalent bonding weakens. These data indicate a large difference in the diradical character between the non-twisted (in blue) and twisted (in orange) molecules and are consistent with those in **Figures 4** and **5**. These data also support the presence of covalent bonding up to around 2.3 Å. There is a region of data points below the trendline from 2.1-2.3 Å that are of interest due to their low $N_{FOD}$ values. These molecules are all derivatives of **2A**, which interestingly all have longer $D_{12}$ distances than their corresponding **1A** derivative counterparts. It is likely that these **2A** derivatives have lower $N_{FOD}$ values because of their simplified "bodies"—which are less conjugated than **1A** derivatives—and thus minimize delocalization of radical electrons. This can be confirmed by visualizing the FOD densities of **1A** and **2A** as seen in **Figure 7**. This figure reveals no FOD density on the simplified "body" of **2A**, as compared to some FOD density on the larger, conjugated "body" of **1A**.

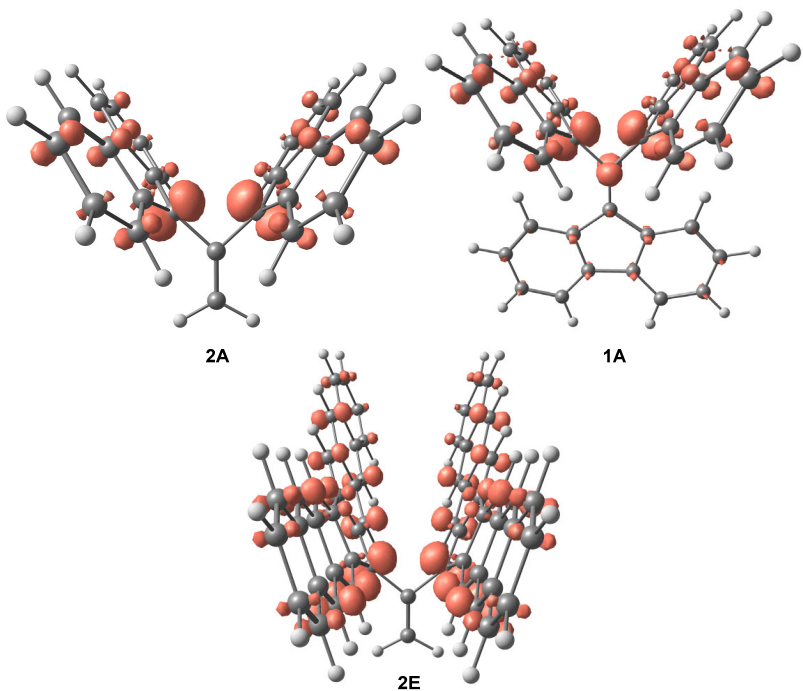

**Figure 7.** FOD density plots for **2A**, **1A**, and **2E** calculated using B3LYP/def2-TZVP model chemistry ($T_e$=9000 K). FOD surfaces are drawn at a 0.005 e au$^{-3}$ level.

There also seems to be an outlier in **Figure 6** with an unusually high $N_{FOD}$ value that refers to molecule **2E**, as indicated by the empty blue circle. Looking at **2E**'s FOD density plot in **Figure 7** reveals a potential reason for its high $N_{FOD}$ value. Compared to the FOD plots of both **1A** and **2A**, it is clear that the radical electrons are significantly more delocalized across the large macrocyle "wings" of **2E**. Since $N_{FOD}$ is calculated by the integration of FOD over all space, $N_{FOD}$ is expected to increase when radical electrons are delocalized over a larger region. Using this reasoning, it would be expected that molecule **2D**, with larger "wings" than **2A** but smaller than those of **2E**, would have an $N_{FOD}$ value between those of **2A** and **2E**. This hypothesis is confirmed by the $N_{FOD}$ values listed in **Table 3**, where for this series of molecules the values increase as follows: 1.37 e, 2.14 e, 3.18 e, for **2A**, **2D**, and **2E**, respectively. As a result of its high FOD value and molecular orbitals seen in **Figure 3**, the interaction between the two wings of molecule **2E** can be described as two pancake-bonded radicals. Since $D_{12}$ in **2E** is too short for pancake bonding, the interaction between C1 and C2 must be covalent in nature. In contrast, the contacts $D_{21,23}$ and $D_{22,24}$ are too long for covalent bonding but within the range for pancake bonding. All of this is to say, **2E** is unlike the rest of the presented molecules in that there is a mix of covalent and pancake bonding, so an increase in its $N_{FOD}$ is to be expected.

For a selected group of molecules listed in **Table 4** two minima were found: one with a non-twisted $C_{2v}$ structure and another twisted $C_2$ structure. While the existence of two isomers was not confirmed for all molecules, we expect that two geometric minima should be present for most of the molecules presented in **Table 1**. In all confirmed cases, however, the twisted $C_2$ conformer was lower in energy to the non-twisted $C_{2v}$ conformer. Since Kubo et al.[1] determined that the higher-energy non-twisted conformer of **1A** was present in the crystal structure, a potential energy scan (PES) was completed to understand the reaction coordinate of such isomerization reactions and why the crystal structure revealed the presence of a higher energy non-twisted isomer. In this work, a relaxed potential energy scan was performed on the simplest molecule in our database, **2A**, to investigate the isomerization reaction pathway between **2A** and **2Atw**. As illustrated in **Scheme 3**, **2A** has a non-twisted $C_{2v}$ isomer (**2A**) and a twisted $C_2$ isomer (**2Atw**). Similar to **1A**, the twisted **2Atw** structure was lower in energy. More specifically, **2Atw**'s ground state energy was 5.59 kcal/mol lower than that of non-twisted **2A**. Since **2A** readily twists into its twisted conformer with slight distortions of $D_{12}$, $D_{12}$ was frozen at each point of the scan to obtain intermediary points along the PES. **Figure 8** shows the isomerization reaction pathway in terms of the molecule's geometry. In the first portion of the figure, as indicated by the blue points below 2.25 Å, torsions $\alpha$ =20-3-1-21 and $\beta$ =20-3-1-22 change little with an increase in $D_{12}$. In this region before 2.25 Å there is no twisting of the "wings". Instead, the central bond weakens through bond elongation while conserving its $C_{2v}$ geometry. At around 2.25 Å, there appear to be two distinct pathways through which the central bond of **2A** breaks: high and low symmetry pathways. The relative energies of each point along these pathways are depicted in **Figure 9**. In the low symmetry pathway, the molecule begins to twist at around 2.25 Å, misaligning the $\pi$ orbitals that make up this elongated $\pi-$ bond, and thus rapidly breaking the central bond and the two above mentioned torsions differ. Then as $D_{12}$ increases with each point after the initial twisting at around 2.25Å up until around 2.50 Å, the molecule relaxes to its **2Atw** conformer. In contrast, in the high symmetry pathway, the $C_{2v}$ geometry is preserved, indicated by the blue points that extend past 2.25 Å. Instead of twisting earlier at around 2.25 Å, the molecule preserves its $C_{2v}$ symmetry where the central bond breaks without twisting by continual elongation of $D_{12}$. At around 2.50 Å, however, the molecule twists into the 2Atw molecule, indicated by the black arrows in **Figures 8** and **9**.

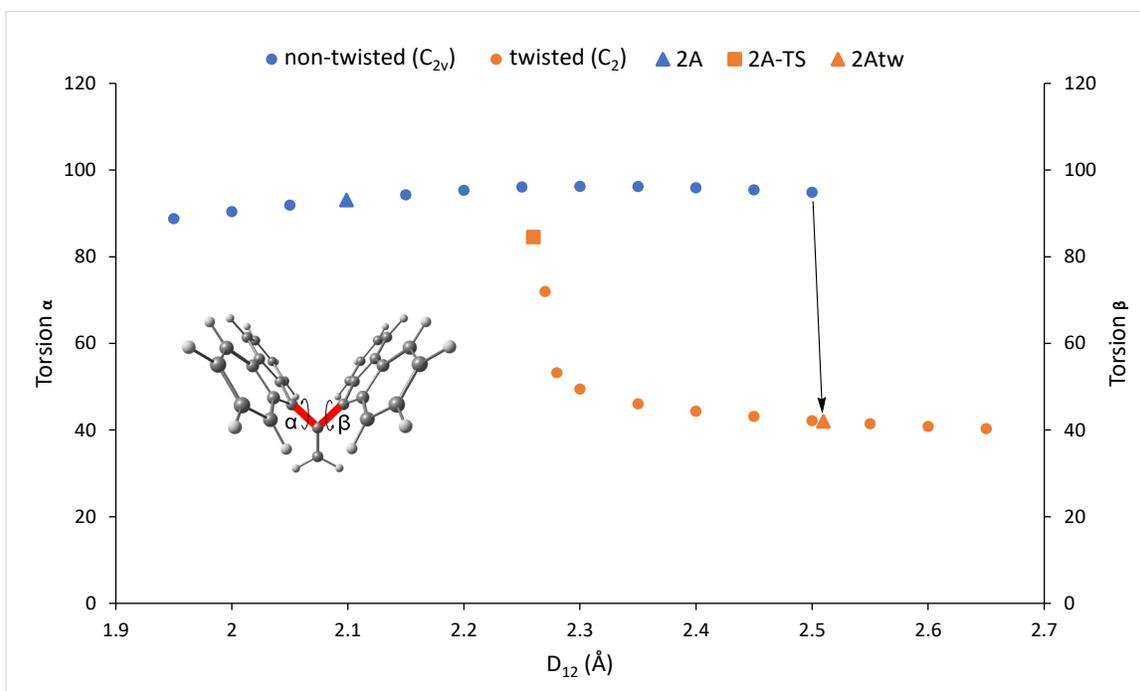

**Figure 8.** Isomerization reaction torsional coordinates along a $D_{12}$ relaxed scan of **2A** comparing torsions α and β and $D_{12}$. The red bonds in the insert indicate the two disrotatory axes of torsion.

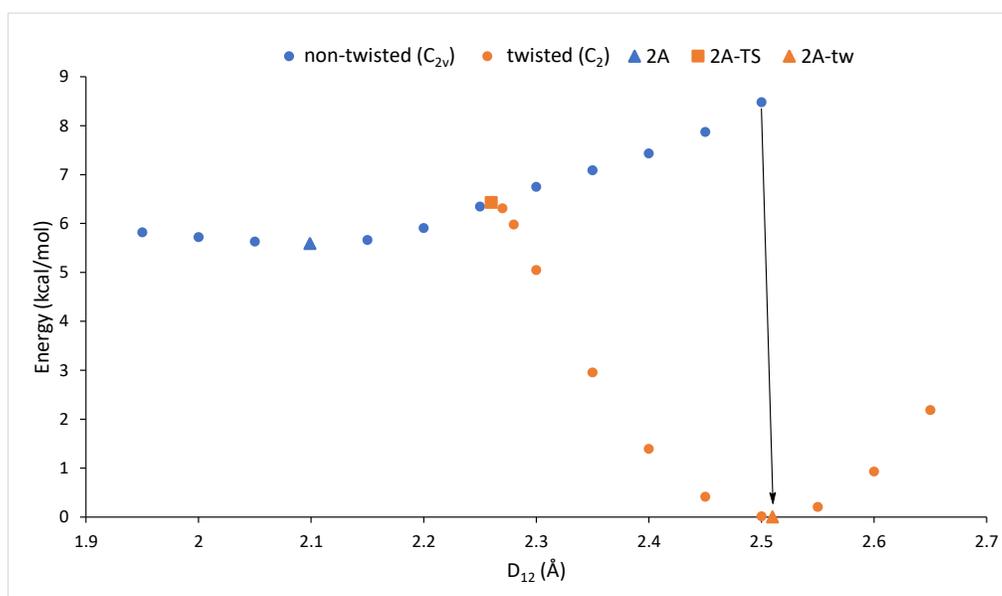

**Figure 9.** Isomerization reaction coordinate plot of 2A comparing energy at each step to $D_{12}$. The point at $D_{12}$=2.5 Å is the longest at which a $C_{2v}$ structure could be optimized. For longer $D_{12}$ values the computations converge to the lower energy twisted $C_2$ structure as indicated by the black arrow.

The misalignment of the π-orbitals after twisting can be seen in **Figure 10** for **2A**. As the "wings" twist, these orbitals no longer overlap well, and thus the central $D_{12}$ bond breaks. **Figure 11** depicts the HOMO molecular orbitals of **2A** along the high and low symmetry pathways from $D_{12}$=2.25 Å to 2.50 Å.

Along the high symmetry pathway, with increasing $D_{12}$, there is less orbital overlap up until 2.50 Å where the bond fully breaks losing electron sharing between the two carbons involved in the central bond. In the low symmetry reaction pathway, the "wings" twist breaking the $C_{2v}$ symmetry and misaligning the π orbitals as early as 2.30 Å. As such, in the low symmetry pathway, the central bond breaks much earlier.

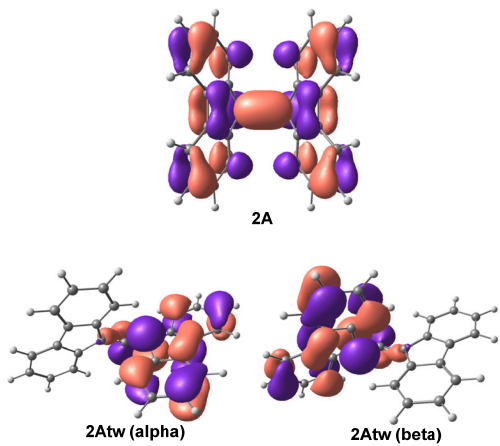

**Figure 10.** HOMO of **2A** and **2Atw** calculated using the UB3LYP-GD3/6-311+G(d,p) method drawn at the 0.03 e au$^{-3}$ level. The π orbitals involved in the central C-C bonding overlap to form a bond in **2A**, however, these orbitals are not aligned for perfect overlap in **2Atw**, preventing orbital overlap and sharing of electrons along $D_{12}$.

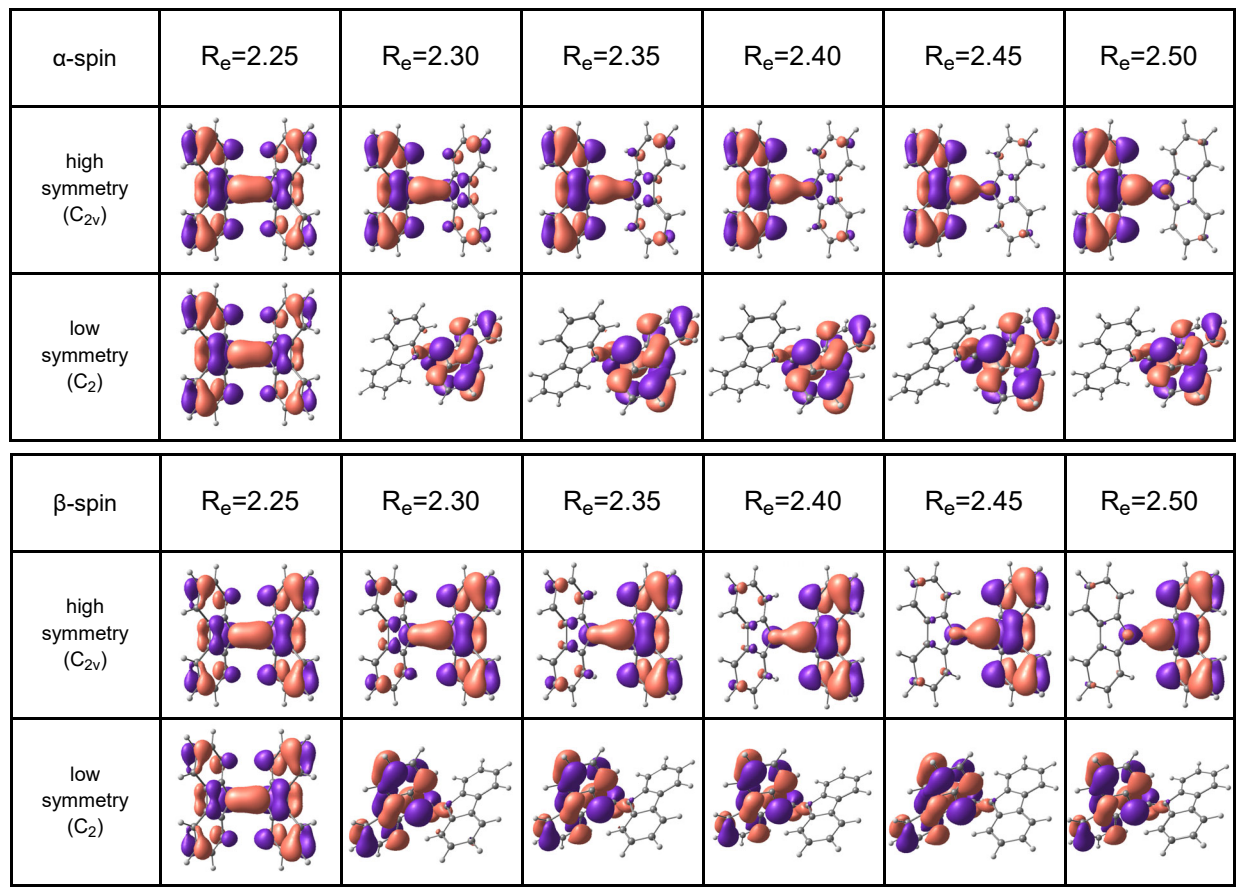

**Figure 11.** High and low symmetry alpha/beta HOMO molecular orbitals of **2A** along PES scan. Molecular orbitals were calculated using the UB3LYP-GD3/6-311+G(d,p) method drawn at the 0.03 e au$^{-3}$ level.

It is expected that the high symmetry pathway is less likely for the isomerization of **2A**. This is because each blue point past 2.25 Å is a high energy conformer that can with any slight deformation adopt a twisted conformation. As seen in **Figure 9**, the activation energy of the isomerization PES is surprisingly low. It had been hypothesized that a large activation energy for such isomerization reactions was the key reason why **1A** had adopted a higher energy conformation in its crystal structure. However, since the activation energy is less than 1 kcal/mol, there must be other effects that restrict molecule **1A** from adopting its lower energy twisted conformer in its crystal structure.

FOD calculations were run on this PES to investigate the diradical character of each conformer. **Figure 12** reveals a large relative increase, from 1.37 e to 2.48 e, in $N_{FOD}$ as **2A** adopts a twisted conformation. While $N_{FOD}$ gradually increases with increasing D$_{12}$, there is a sharp increase starting at around 2.25 Å as the molecule twists. This indicates that diradical character significantly increases as the "wings" of **2A** twist and the central bond breaks. After the bond has broken, further increases in bond length from 2.30 to 2.65 Å have little effect on the diradical character. This spike in $N_{FOD}$ indicates the presence of a covalent C-C bond for **2A** before any twisting takes place.

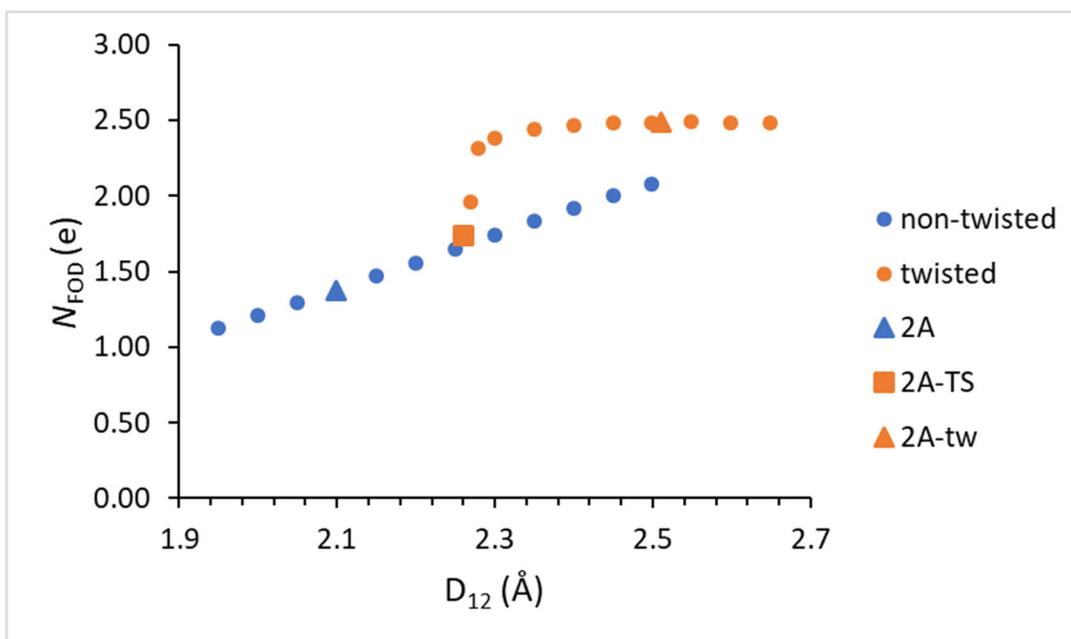

**Figure 12.** Isomerization reaction scan of $N_{FOD}$ for **2A** at B3LYP/def2-TZVP ($T_e$=9000 K) level of theory.

**Table 4.** Physical parameters of molecules that exhibit a lower energy twisted conformer relative to the untwisted conformer at the UB3LYP-GD3/6-311+G(d,p) level of theory.

| Molecule | $R_e$ (Å) | WBI | $\Delta BDE_{isomers}$ (kcal/mol) | $\Delta BDE_{ST}$ (kcal/mol) | Diradical Character ($y_0$) |
|---|---|---|---|---|---|
| 1Atw | 2.466 | 0.147 | 3.63 | -2.76 | 0.895 |
| 1A(Br:4)tw | 2.479 | 0.017 | -[a] | -2.02 | 0.963 |
| 2Atw | 2.510 | 0.018 | 5.59 | -1.48 | 0.736 |
| 2A(Me:4,11,12,19)tw | 2.604 | 0.016 | 7.14 | -0.924 | 0.614 |
| 2A(Br:4,11,12,19)-tw | 2.610 | 0.029 | 6.54 | -0.678 | 0.550 |
| 2Dtw | 2.510 | 0.022 | 3.15 | -1.29 | 0.824 |
| 10Atw | 2.481 | 0.107 | 7.21 | -1.00 | 0.438 |
| 10A(Me:4,12)tw | 2.490 | 0.014 | -[a] | -1.19 | 0.555 |

[a]There is only a twisted minimum

The isomerization reaction involving twisting of the "wings", as depicted in **Scheme 3**, was investigated for selected molecules seen in **Table 4**. All equilibrium bond distances were near ~2.5 Å, which is the limit predicted by both WBI and $BDE_{ST}$ correlations where no C-C WBI or bond dissociation energy is expected. Furthermore, low WBI and high diradical character suggest each twisted molecule is in its diradical state without the presence of a $D_{12}$ bond. It should also be noted that the $\Delta BDE_{isomers}$ values are positive for all molecules where both a twisted and non-twisted conformer was present, indicating that the twisted triplet (diradical) conformation is lower in energy than the non-twisted singlet molecule. Since the twisted isomer was found to always be lower in energy and the isomerization reaction of **2A** revealed a small activation energy, crystal packing effects were investigated as a possible stabilizing effect for the higher energy non-twisted conformer. In fact, Kubo et al. suggested[1] that **1A** adopts the untwisted $C_{2v}$ conformation as a result of crystal packing effects where the two fluorenyl rings face each other in a perpendicular configuration.[1] It should be noted that for molecules **1A(Br:4)tw** and **10A(Me:4,12)tw**, no non-twisted conformer was found. For these molecules the steric repulsions due to the sizes of the halogen atom or two methyl groups were too large for non-twisted energy minima to exist. To minimize steric hinderances, these molecules are forced to adopt their twisted conformation. This means that there is a limit to the size and number of substituents one can place on the "wings" of the target molecules to further elongate $D_{12}$.

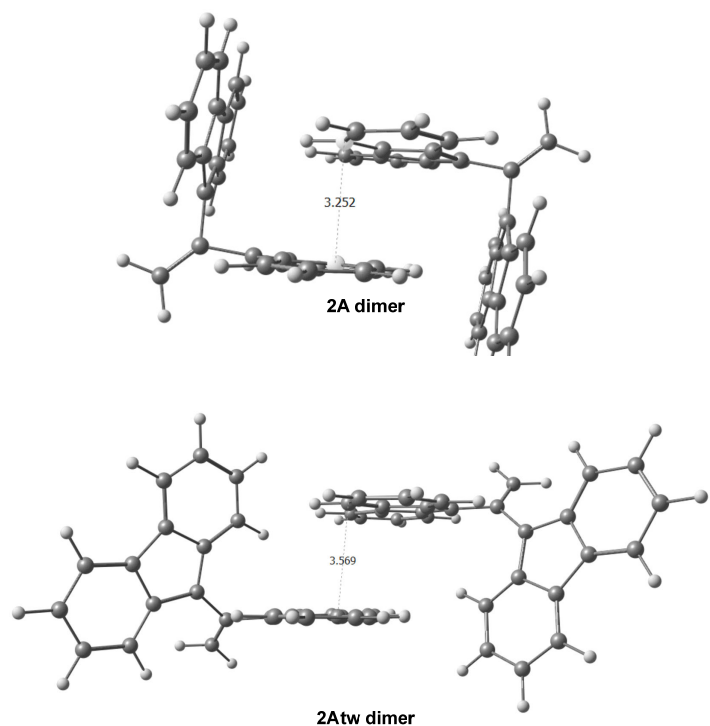

**Figure 13.** Optimized dimer structures of **2A** and **2Atw**. The closest C-C distances between the wings of each dimer are displayed.

Dimer geometry calculations were completed for **2A**, one of the simplest target molecules, to investigate non-bonding crystal packing effects, as described by Kubo et al.[1] **Figure 13** illustrates the packing for the dimers of **2A** and **2Atw**. For the **2A** dimer, the "wings" of the neighboring monomer appear to lock each molecule of the dimer in its non-twisted form. Unlike in the monomer, where there is space for the "wings" to twist, as a dimer, this space is taken up by the opposing molecule, restricting the twisting isomerization reaction from taking place. While steric repulsions likely play a significant role in stabilizing the **2A** dimer in the non-twisted conformation, there are also non-bonding interactions that further stabilize the **2A** dimer. In the non-twisted dimer, the "wings" of each molecule are more closely packed and overlap more compared to the **2Atw** dimer (**Figure 13**). This close packing results in a larger non-bonding vdW interaction energy. In fact, this interaction energy for the **2A** dimer (-23.6 kcal/mol) is almost twice as large as for the twisted dimer (-12.8 kcal/mol). When considering this large stabilizing energy for the non-twisted dimer, **2A** would likely remain in its non-twisted form in its crystal structure despite the twisted monomer being a lower energy conformation and the low activation energy of the isomerization reaction. This finding supports Kubo's observation that crystal packing effects stabilize **1A** in its bonded, non-twisted form. While in-depth analysis for the isomerization reactions and packing was not completed for all the target molecules, these preliminary findings suggest that for all the molecules that exhibit lower-energy twisted isomers, the non-twisted conformation would be preferred in their crystal structure. The stabilization resulting from non-bonding vdW's interactions appears to be significant enough to favor the non-twisted bonded conformation of these target molecules.

$^{13}$C NMR spectroscopy is a sensitive tool to explore the hybridization and environment of carbon atoms. **Figure 14** displays the computed $^{13}$C NMR chemical shifts for **2A** in the bonded ($C_{2v}$) and twisted diradicaloid ($C_2$) conformation. According to the calculation, the peak around 100 ppm corresponds to the chemical shifts of C1 and C2 in the bonded conformation, while this peak moves by ~50 ppm to a much higher value when the bond is broken (**2Atw**). A similar major increase in chemical shift is seen for pairs **1A/1Atw**, **2D/2DTw**, and **10A/10Atw** as shown in **Figures S4**, **S5**, and **S6**, respectively. **Figure S7** shows the development of a similar shift by almost 100 ppm as the single bond is gradually broken in ethane. It appears that $^{13}$C NMR spectroscopy offers a tool to monitor these extremely elongated C-C single bonds.

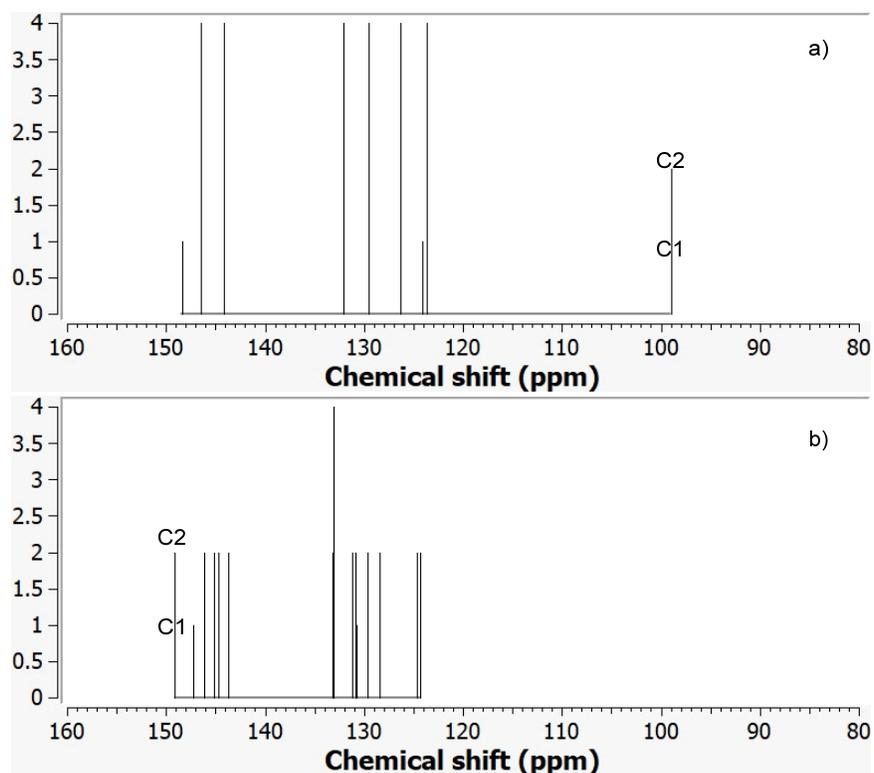

**Figure 14.** Theoretically predicted $^{13}$C chemical shifts of a) **2A** and b) **2Atw** calculated by GIAO-B3LYP-GD3/6-311+G(d,p) method. The structures were optimized at the UB3LYP-GD3/6-311+G(d,p) level of theory and converged to a $C_{2v}$ symmetry for **2A** and $C_2$ symmetry for **2Atw**. TSM was computed at B3LYP-GD3/6-311+G(d,p) and used as the reference.

**Conclusions**

Generally, it is assumed that a "forbidden zone" exists separating the extremely elongated single C-C bond distances from the shortest of the intermolecular pancake bonds as illustrated in **Figure 1**. The discovery by Kubo et al. has reduced this forbidden range by increasing the lower limit to about 2.04 Å. The presented analysis of a wide-ranging selection of molecules and molecular models predicts that the upper limit of extremely stretched C-C single bonds should be revised to about 2.2 Å. The trends described in this work show that the strengths of the extremely stretched C-C bonds decrease nearly linearly with increasing bond length, parallel with the decrease of the computed WBI values. Importantly, no diradical character was exhibited in the designed target molecules with computed equilibrium bond lengths exceeding 2.0 Å. Only in molecules that adopt a twisted configuration was the bond broken, creating a diradical. Finally, we highlight that $^{13}$C NMR chemical shift values depend sensitively on the length of the extremely elongated C-C bond providing potentially a tool for their characterization.[1]

**Supporting Information** contains figures of atomic numbering, predicted Raman and 13C NMR spectra, details of DFT calculations.

**Acknowledgements**

Support by the U. S. National Science Foundation for this research (M.K. grant number CHE-2107820 and H.L. grant number CHE-2107820) is gratefully acknowledged. M. K. and E. J. J. K. are grateful for ample computer time provided by the High-Performance Computer Center, Georgetown University. H.L. and J.C. are grateful for ample computer time provided by the High-Performance Computer Center of the Texas Tech University.

TOC Graphic

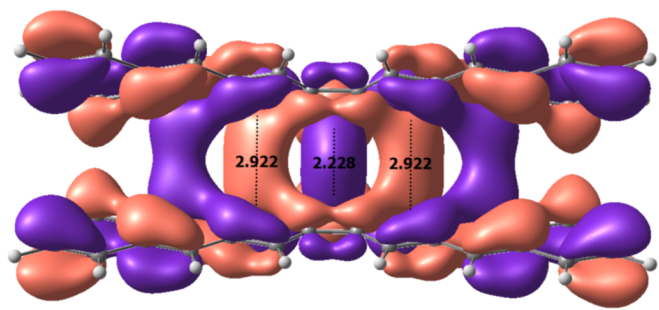

# Correction to "Extremely Long C−C Bonds Predicted beyond 2.0 Å"

Eero J. J. Korpela, Jhonatas R. Carvalho, Hans Lischka, and Miklos Kertesz*

J. Phys. Chem. A 2023, 127, 20, 4440–4454.    DOI: 10.1021/acs.jpca.3c01209

This is a correction of the grant no. for the support by the U.S. National Science Foundation for Hans Lischka. The correct grant no. is CHE-2107923.